\documentclass[aps,prd,amsmath,amssymb]{revtex4}

\usepackage{bm}
\usepackage{amsmath}
\usepackage{amssymb}
\usepackage{latexsym}
\usepackage{amsfonts}
\usepackage{epsfig}
\usepackage{psfrag}
\usepackage{graphicx}

\newcommand{\ord}{{\cal O}}

\newcommand{\ket}[1]{\left|#1\right>} 
\newcommand{\bra}[1]{\left<#1\right|}

\newcommand{\f}[1]{\mbox{\boldmath$#1$}}

\newcommand{\na}{\mbox{\boldmath$\nabla$}}
\newcommand{\bea}{\begin{eqnarray}}
\newcommand{\ea}{\end{eqnarray}}

\begin{document}

\title{Fundamental Quantum Effects from a Quantum-Optics Perspective}

\author{Ralf Sch\"utzhold}

\affiliation{
Fakult\"at f\"ur Physik, Universit\"at Duisburg-Essen,\\ 
D-47048 Duisburg, Germany\\
E-mail: ralf.schuetzhold@uni-due.de}

\begin{abstract}
This article provides a brief overview of some fundamental effects of 
quantum fields under extreme conditions.
For the Schwinger mechanism, Hawking radiation, and the Unruh effect,
analogies to quantum optics are discussed, which might help to approach 
to these phenomena from an experimental point of view. 
\end{abstract}



\maketitle

\section{Introduction}\label{intro}

In quantum field theory, the vacuum state id not just empty space, 
but a complicated state filled with quantum fluctuations. 
If we apply some extreme conditions such as a strong electric field,
these vacuum fluctuations may manifest themselves in the creation 
of real particle pairs -- which is a pure quantum effect.
There are several examples for such phenomena -- 
in the following, we shall focus on three cases: 

\vspace{0.5cm}
\begin{tabular}{|c|c|c|}
\hline
Fundamental effect & Extreme condition & Experimental approach? \\
\hline
Schwinger mechanism & electric field & ultra-strong laser field \\
Hawking radiation & gravitational field & black hole analogues \\
Unruh effect & acceleration & electrons in laser field \\
\hline
\end{tabular}
\vspace{0.5cm}

In all of these examples, it will turn out to be interesting to apply 
ideas from quantum optics.
Since none of these effects has been (directly) observed yet, 
it is also very interesting and desirable to find an experimental 
approach.  

\section{Schwinger mechanism}\label{Schwinger mechanism}

\subsection{Dirac sea}

Before discussing the Schwinger mechanism, let us briefly review the 
concept of the Dirac sea: 
As is well known, identifying $i\hbar\partial_t$ with the energy
$\cal E$ and $-\hbar^2\na^2$ with the momentum squared $p^2$, the 
Schr\"odinger equation yields the non-relativistic energy momentum 
relation
\bea
i\hbar\frac{\partial}{\partial t}\psi
=
-\frac{\hbar^2}{2m}\,\na^2\psi+V\psi
\quad
\leadsto
\quad
{\cal E}=\frac{p^2}{2m}+V
\,.
\ea
The correct relativistic description of electrons, for example, 
is given by the Dirac equation\cite{Dirac}
\bea
\gamma^\mu
\left(
i\hbar\partial_\mu+qA_\mu\right)\Psi
=mc\Psi
\;
\leadsto
\;
{\cal E}=V\pm\sqrt{c^2p^2+m^2c^4}
\,,
\ea
which reproduces the relativistic energy momentum relation.
However, this relation -- and also the Dirac equation -- 
always has positive and negative energy solutions, as 
indicated by the $\pm$ sign in front of the square root above.
In order to facilitate stable electron solutions at positive 
energies, the negative energy levels are completely filled in 
the vacuum state according to the Dirac sea picture, see 
Fig.~\ref{dirac-schwinger}a. 
In this way, the Pauli principle prevents the decay of electrons 
into the negative energy levels.
Holes in the Dirac sea then correspond to positrons, whose 
existence was in this way foreseen by Dirac even before their 
experimental discovery. 

\subsection{Tunneling}

\begin{figure}[hbt]
\epsfig{file=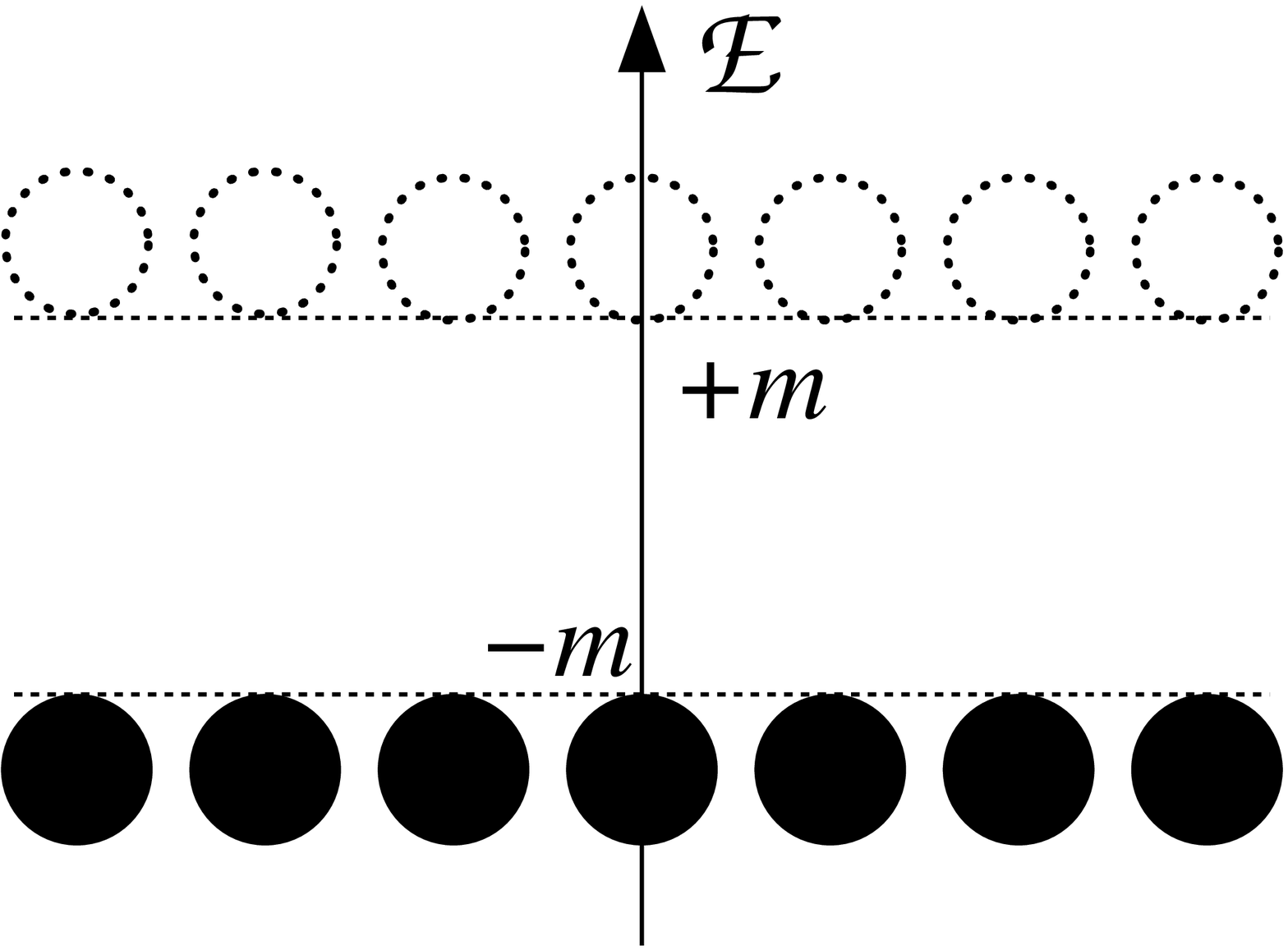,width=.4\textwidth}
\hfill
\epsfig{file=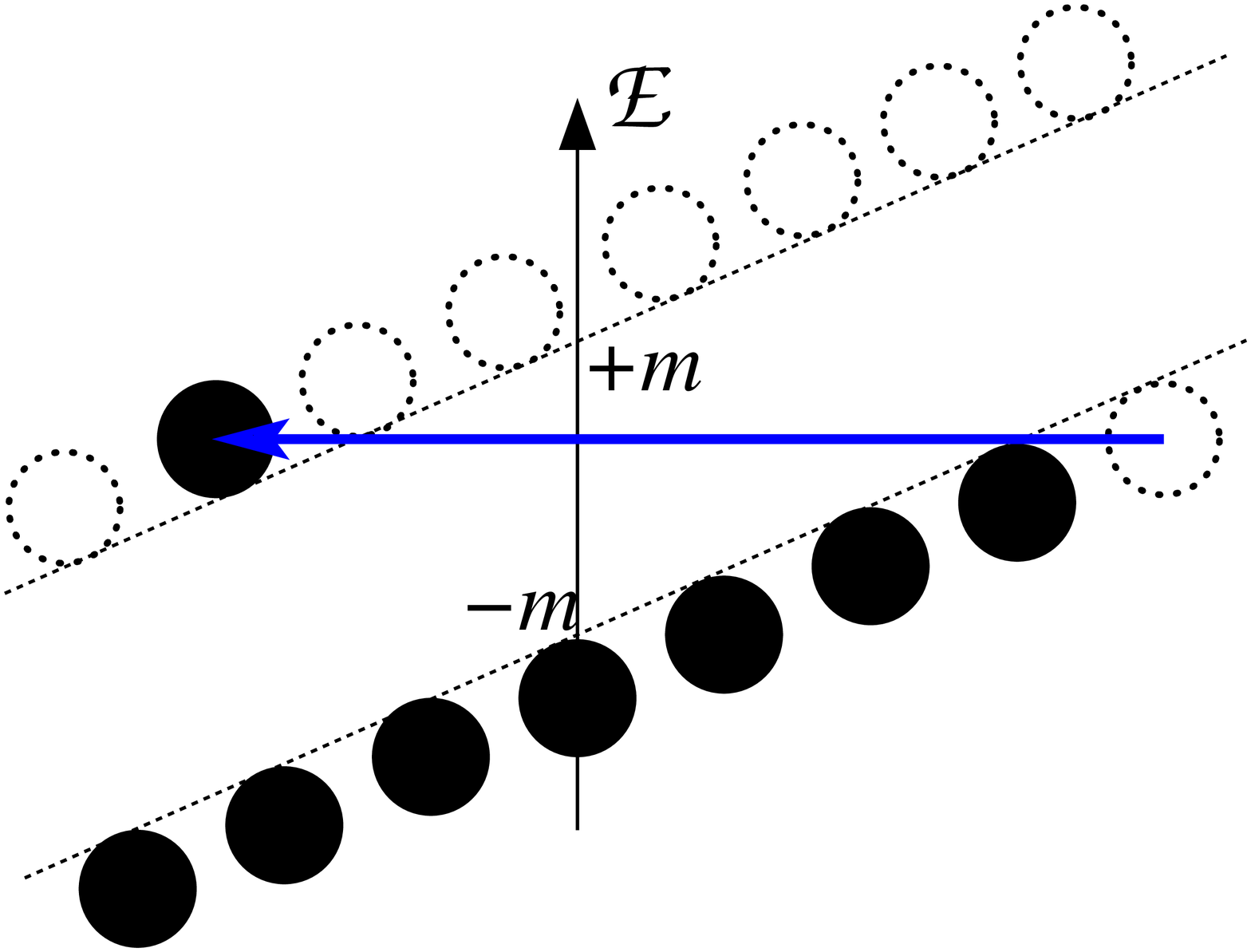,width=.5\textwidth}
\caption{Sketch of the Dirac sea in vacuum (left) and with a constant 
electric field (right) where an electron can tunnel from the Dirac sea 
to the positive energy levels (Schwinger mechanism).}
\label{dirac-schwinger}
\end{figure}

Now let us see what happens to the Dirac sea in the presence of a 
constant electric field $\f{E}=E\f{e}_x$.
In this case, we get an additional potential $V(x)=qEx$, which 
tilts the level spectrum, see Fig.~\ref{dirac-schwinger}b. 
As a result, it becomes possible that an electron in the Dirac sea 
has the same energy as an empty positive energy level on the 
other side of the gap. 
Classically, the gap (between $-mc^2$ and $+mc^2$) is forbidden by 
energy arguments -- but is quantum theory, particles may tunnel 
through this region.

Such a tunneling process corresponds to the creation of an 
$e^+e^-$ pair out of the vacuum due to the electric field.
Let us estimate the probability for such an event.
In quantum mechanics, tunneling is exponentially suppressed.
The exponent can be estimated by the length $L$ of the tunneling
barrier times the corresponding potential difference $\Delta V$ 
(in the relativistic case).
For a constant electric field $E$, the length can be calculated 
easily by energy conservation $qEL=2mc^2$.
The height of the potential barrier, however, is not constant 
and thus more complicated -- but we may get a rough estimate by 
setting $\Delta V=\ord(mc^2)$.
In this way, we get the following rough estimate for the 
$e^+e^-$ pair creation probability 
\bea
P_{e^+e^-}
\propto
\exp\left\{-\frac{L\Delta V}{\hbar c}\right\} 
\propto
\exp\left\{-\frac{2mc^2}{qE}\,\frac{\ord(mc^2)}{\hbar c}\right\} 
\,.
\ea
This result is already rather close to the truth -- an 
exact calculation yields the tunneling 
exponent\cite{Sauter,Heisenberg+Euler,Schwinger}
\bea
\label{exponent}
P_{e^+e^-}
\propto
\exp\left\{-\pi\,\frac{c^3}{\hbar}\,\frac{m^2}{qE}\right\} 
=
\exp\left\{-\pi\,\frac{E_S}{E}\right\} 
\,,
\ea
where we have introduced the Schwinger critical field strength 
\bea
E_S=\frac{c^3}{\hbar}\,\frac{m^2}{q}\approx1.3\times10^{18}\,{\rm V/m}
\,.
\ea
Apart from $e^+e^-$ pair creation, this field strength does also set 
the scale where the QED vacuum starts to behave as a non-trivial 
medium and shows effects such as birefringence etc.
This critical field strength corresponds to an intensity of 
$I_S=\ord(10^{29}\rm W/cm^2)$.
Such intensities are currently beyond our experimental 
capabilities.
Planned ultra-strong lasers\cite{ELI}
have an envisioned peak intensity of $I=\ord(10^{26}\rm W/cm^2)$.
Inserting the maximum electric field achievable with these lasers,
we obtain an exponential suppression of the 
$e^+e^-$ pair creation probability of 
$\exp\{-\pi E_S/E\}=\ord(10^{-61})$.
Unfortunately, this number is too small to be detectable. 
Therefore, we are led to the question of whether one might 
enhance this probability. 

\subsection{Assisted Tunneling}

As we have discussed in the previous Section, the laser intensities 
envisioned in the near future are probably not large enough to observe 
the Schwinger mechanism directly.
Therefore, we are led to the question of whether (and how) one could 
enhance this effect.
To achieve this goal, we borrow an idea known from quantum optics --
assisted tunneling.
Schwinger pair creation is similar to the ionization of an H-atom,
for example, by a constant electric field. 
This well-know process can be enhanced by sending additional 
electromagnetic waves to the atom.
Applying the same idea to the Schwinger mechanism, we studied the 
$e^+ e^-$ pair creation is a constant electric field which is 
superimposed by a plane wave x-ray beam. 
As one would expect from the analogy to quantum optics, 
the x-ray beam can lead to an enhanced pair creation probability 
by reducing the tunneling exponent\cite{Dunne}. 

\begin{figure}[hbt]
\epsfig{file=schwinger.eps,width=.47\textwidth}
\hfill
\epsfig{file=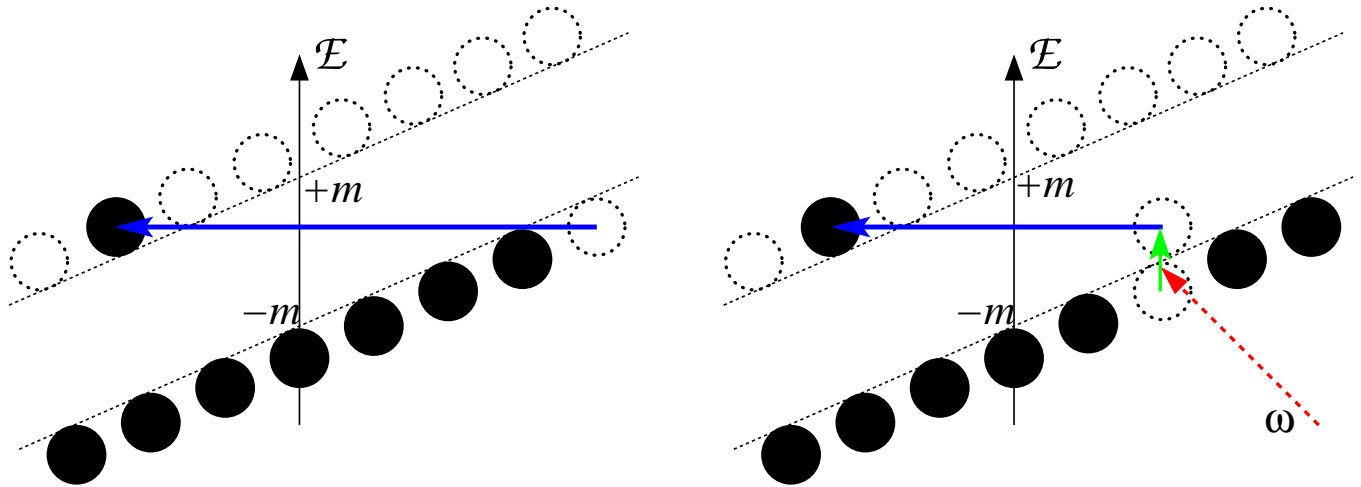,width=.47\textwidth}
\caption{Sketch of the Schwinger mechanism with a constant electric 
field (left) and an additional x-ray beam of frequency $\omega$ (right).
The x-ray helps the electron to penetrate the classically forbidden 
region (mass gap $2m$) such that the remaining way to tunnel is shorter.}
\label{schwinger-assist}
\end{figure}

As an intuitive picture, one can imagine that the x-ray provides 
some additional energy and so helps the electron to penetrate the 
gap a bit -- such that the remaining way to tunnel is reduced, 
see Fig.~\ref{schwinger-assist}.
The maximum enhancement we could achieve -- while still preserving 
the non-perturbative nature of the process -- is given by\cite{catalysis}
\bea
P_{e^+e^-}
\propto
\exp\left\{-\pi\,\frac{E_S}{E}\right\} 
\to
\exp\left\{-(\pi-2)\frac{E_S}{E}\right\} 
\,.
\ea
Inserting the same values as in the previous Section, we find that 
the exponential suppression is no longer $10^{-61}$
(as in the case without x-rays) but now $10^{-22}$.
This drastic enhancement might facilitate an experimental realization.

\subsection{So What?}

In view of the experimental difficulties mentioned above, one could ask the 
question of why we should try to observe the Schwinger mechanism. 
After all, $e^+ e^-$ pair creation by colliding gamma rays has be observed 
already. 
The main reason for our interest in the Schwinger mechanism lies in its 
purely non-perturbative nature.
The majority of QED effects can be understood from Feynman diagrams 
which are based on perturbation theory.
In this approach, the S-matrix, for example, is Taylor expanded in powers 
of the coupling (i.e., charge $q$)
\bea
\bra{{\rm out}}\hat S\ket{{\rm in}}=
a_0+a_1q+a_2q^2+\dots  
\ea
In contrast, the Schwinger mechanism is a purely non-perturbative 
QED vacuum effect.
This can already be seen from Eq.~(\ref{exponent}) 
\bea
P_{e^+e^-}
\propto
\exp\left\{-\pi\,\frac{c^3}{\hbar}\,\frac{m^2}{qE}\right\} 
=
\exp\left\{-\pi\,\frac{E_S}{E}\right\} 
\,,
\ea
which does not admit a Taylor expansion in $q$ (nor $E$), i.e., we have 
an essential singularity at $q=0$. 
Therefore, no Feynman diagram can ever describe the Schwinger effect. 
In quantum chromo-dynamics (QCD), such non-perturbative effects are 
very important for describing experimental data, but direct experimental 
tests are complicated by additional properties (e.g., confinement).
Thus, quantum electrodynamics offers the possibility of performing
more controlled experiments -- but non-perturbative QED vacuum effects
such as the Schwinger mechanism have not been observed yet. 


\section{Hawking Radiation}\label{Hawking Radiation}

\subsection{Black Hole Evaporation}

In close analogy to the strong electric field which rips apart vacuum 
fluctuations and so produces $e^+ e^-$ pairs via the Schwinger mechanism,
a strong gravitational field can also rip apart vacuum fluctuations
leading to the creation of particle pairs.
A prominent example is Hawking radiation as a result of the distortion 
of quantum vacuum fluctuations due to event horizon of a black hole.
At the horizon, all wave-packets are ripped apart in the course of 
time -- the part inside is trapped and finally hits the singularity 
whereas the part of the wave packet outside the horizon can escape 
to infinity.
Assuming the the wave packet was initially in its ground state (vacuum),  
this process of ripping it apart is a drastic departure from equilibrium 
and thus results in an excitation. 
It turns out that the so generated occupation of the outgoing part 
of the wave packet is thermal with the Hawking temperature\cite{Hawking}
\bea
T_{\rm Hawking}
=
\frac{1}{8\pi M}
\frac{\hbar\,c^3}{G_{\rm N}k_{\rm B}}
\,.
\ea
Even though the above formula for the Hawking temperature is quite 
simple, it combines four (apparently) different areas of physics via 
the occurring natural constants:
quantum theory ($\hbar$), relativity ($c$), gravity ($G_{\rm N}$),  
and thermodynamics ($k_{\rm B}$). 
It almost seems as if nature is trying to give us a hint. 
Indeed, understanding the origin of the unexpected thermal nature 
of black holes (black hole entropy etc.) will probably be a big step 
towards finding the correct laws of nature that unify gravity and 
quantum theory. 
Unfortunately, for typical astronomical black hole swith a mass of,
say, 30 solar masses, the Hawking temperature is around 2~nK and thus 
probably not observable. 

\subsection{Black Hole Analogues}

The contrast between the fundamental importance of Hawking radiation 
and the difficulty of observing it motivates alternative approaches.
Interestingly, it turns out that phonons propagating in fluids satisfy 
the same equation of motion as quantum fields in curved 
space-times\cite{Unruh-prl}. 
Then, remembering the famous quote of R.~Feynman 
{\em ``The same equations have the same solutions.''}, we conclude that 
it should (in principle) be possible to achieve the analogue of 
Hawking radiation in the laboratory.
Indeed, a de~Laval Nozzle is analogous to a black hole:
At the entrance of the nozzle, the flow is sub-sonic and thus 
phonons can propagate in all directions.
This region is analogous to the exterior of the black hole. 
When flowing through the nozzle, the fluid speeds up and exits it 
with super-sonic flow velocity.
In this region, all phonons are dragged away by the flow and thus 
cannot propagate upstream anymore.
Therefore, this is the analogue of the black hole interior where 
everything is trapped.
The border between these two regions lies at the most narrow point 
of the nozzle, where the flow velocity exceeds the speed of sound.
Consequently, this border is analogous to the event horizon.
Repeating Hawking's derivation of black hole evaporation for this 
system, we find that the nozzle emits thermal phonons with the 
temperature\cite{Unruh-prl}
\bea
\label{analogue-Hawking}
T_{\rm Hawking}
=
\frac{\hbar}{2\pi\,k_{\rm B}}\,
\left|\frac{\partial}{\partial r}\left(v_0-c_{\rm s}\right)\right|
\,,
\ea
which is determined by the velocity gradient at the horizon. 
Depending on the experimental realization, this temperature could 
range from some nano-Kelvin (for Bose-Einstein condensates) up to 
fractions of a Kelvin.

It should be mentioned here that the analogy to gravity is not 
restricted to phonons, but applies under certain conditions to 
other quasi-particles in condensed matter\cite{LNP}.
The described analogy cab be used in basically two ways:
First, as a toy model for the underlying theory 
(including quantum gravity).
For example, one can study  the dependence of Hawking radiation 
on modifications of the dispersion relation at large energies\cite{origin}. 
%
Second, it would be very nice to do experiments and actually measure 
(the analogue of) Hawking radiation in the laboratory. 

\subsection{Detectability?}

As one may infer from Eq.~(\ref{analogue-Hawking}), there are basically 
two ways to make facilitate the detection of (the analogue of) Hawking 
radiation in the laboratory.
First, one could try to achieve large velocities (and thus large gradients) 
to increase the Hawking temperature in Eq.~(\ref{analogue-Hawking}). 
This route is taken in optical or electromagnetic (wave-guide) analogues 
of black holes\cite{wave-guide,Philbin}. 
The second route is to achieve high measurement accuracy in order to 
be able to detect low temperatures.
This is particularly necessary for phonons in Bose-Einstein 
condensates\cite{Garay} with a sound speed of order mm/s. 
Let us focus on the second case and discuss how to measure a few phonons 
with small energies $k_{\rm B}T=\ord(10^{-13}\,{\rm eV})$ and a broad 
(thermal) spectrum.
One idea is to use doubly detuned optical Raman transitions\cite{Raizen}.
To this end, we assume that the atoms forming the Bose-Einstein condensate
possess an internal electronic structure consisting of two meta-stable 
(ground) states plus some excited state(s).
Now, with two laser beams, we can drive Raman transitions between the 
two meta-stable states. 
However, if we detune one of the laser frequencies a little bit, these
transitions are no longer possible due to energy conservation -- 
if the Bose-Einstein condensate is in its ground state.
But in the presence of phonon excitations, the transition can be possible 
if the missing energy (detuning) of, say $\delta=\ord(10^{-13}\,{\rm eV})$
is compensated by the simultaneous absorption of a phonon with this 
(or a higher) energy.
In this way, single phonons are transformed into single atoms in the 
second meta-stable state.
Fortunately, separating atoms in different electronic states is possible 
(optical tweezers) and counting single atoms is possible with present-day
technology. 
In summary, observing (the analogue of) Hawking radiation in the 
laboratory. is certainly an experimental challenge, 
but not completely impossible.  

\subsection{Hints for Quantum Gravity?}

As mentioned before, the Hawking effect is very important from a fundamental 
point of view.
Let us discuss this aspect in some more detail.
If we start from the formula for the Hawking temperature 
\bea
T_{\rm Hawking}
=
\frac{1}{8\pi M}
\frac{\hbar\,c^3}{G_{\rm N}k_{\rm B}}
\,,
\ea
and try to construct an analogue to the first law of thermodynamics
(which is basically just energy conservation) 
\bea
dE=dMc^2=T_{\rm Hawking}\,dS_{\rm BH}+\dots
\,,
\ea
we obtain the black hole (Bekenstein) entropy\cite{thermo} 
\bea
S_{\rm BH}
=
\frac{A}{4}\,\frac{k_{\rm B}c^3}{G_{\rm N}\hbar}
=
\frac{A}{4}\,\frac{k_{\rm B}}{\ell_{\rm Planck}^2}
\,.
%
\ea
Apart from Boltzmann's constant $k_{\rm B}$, the entropy is given by 
a quarter of the horizon area $A$ in units of the Planck length squared.
This length scale obtained from the natural constants $c$, $G_{\rm N}$, 
and $\hbar$ is believed to be a characteristic scale for quantum gravity.
That's why many scientists believe that truly understanding the Hawking 
effect will help us to solve puzzles of quantum gravity such as the 
black hole information ``paradox''. 

\section{Unruh Effect}\label{Unruh Effect}

\subsection{Principle of Equivalence}

As our final example, let us discuss the Unruh effect.
Interestingly, the are deep connections between Hawking radiation and 
the Unruh effect: 
Similar to Einstein's gedanken experiments with elevators, we may 
compare a stationary observer near a black hole and a freely falling 
observer on the one hand with an accelerated observer and an 
inertial observer in flat space-time on the other hand.
The stationary observer at a fixed distance to the black hole 
feels the gravitational attraction and is -- via the equivalence 
principle -- locally equivalent to an accelerated observer in 
flat space-time.
Similarly, the freely falling observer is locally equivalent to 
an inertial observer in flat space-time.
As one would expect from this analogy, the former two observers see a 
thermal spectrum whereas the latter two do not observe particles.
In flat space-time, this is the essence of the Unruh effect\cite{Unruh-prd}.

\subsection{Accelerated Observer}

The Unruh effect states that a uniformly accelerated detector 
(in flat space-time) experiences the inertial (Minkowski) vacuum state 
as a thermal bath with the Unruh temperature\cite{Unruh-prd} 
\bea
T_{\rm Unruh}
=
\frac{\hbar}{2\pi k_{\rm B}c}\,a
=
\frac{\hbar c}{2\pi k_{\rm B}}
\,\frac{1}{d_{\rm horizon}}
\,,
\ea
where $a$ is the acceleration. 
Such a uniformly accelerated observer cannot see (nor send signals to)
the full space-time -- it is causally disconnected with a whole region 
(the Rindler wedge).
The minimum spatial distance between the observer and this hidden region 
is given by the horizon length $d_{\rm horizon}=c^2/a$.
For everyday accelerations such as $a=9.81\,{\rm m/s^2}$, the Unruh 
temperature is very small $T_{\rm Unruh}\approx 4\times10^{-20}\,{\rm K}$
and thus probably not observable. 
However, electron is a strong laser field, for example, experience 
much larger accelerations -- which might lead to observable 
signatures\cite{Chen+Tajima}.

\subsection{Accelerated Electrons}

Now, let us consider an electron accelerated by a laser field.
For simplicity, we assume the acceleration to be constant.
Of course, this is not quite correct, but the qualitative picture 
remains the same.
An observer co-moving with the accelerated electron would see a 
thermal bath of photons.
Since the electron possess a finite cross section due to Thomson
(or Compton) scattering, the observer would conclude that the electron 
scatters a photon out of the thermal bath into another mode with a 
finite probability (per unit time).
Such a scattering event in the accelerated frame -- after translation 
back into inertial (laboratory) frame -- 
corresponds\cite{Happens,Habs}
to the emission of a real photon {\em pair}. 
Thus, we get a conversion of (virtual) quantum vacuum fluctuations 
into (real) particle {\em pairs} by non-inertial scattering.  
This is a pure quantum effect, which {\em cannot} be explained 
within classical electrodynamics.
For example, the emitted photon pairs are entangled 
(in energy and  polarization).
Thus, they can be understood as signatures of the Unruh Effect. 

\subsection{Analogy to Quantum Optics}

For the signatures of the Unruh effect discussed above, there are also 
interesting analogies to quantum optics.
Spontaneous parametric down-conversion is a very important process in 
this field. 
It is basically the main source for entangled photon pairs, which can be 
used for many interesting applications, such as teleportation, 
quantum cryptography, etc.  

\begin{figure}[hbt]
\center
\epsfig{file=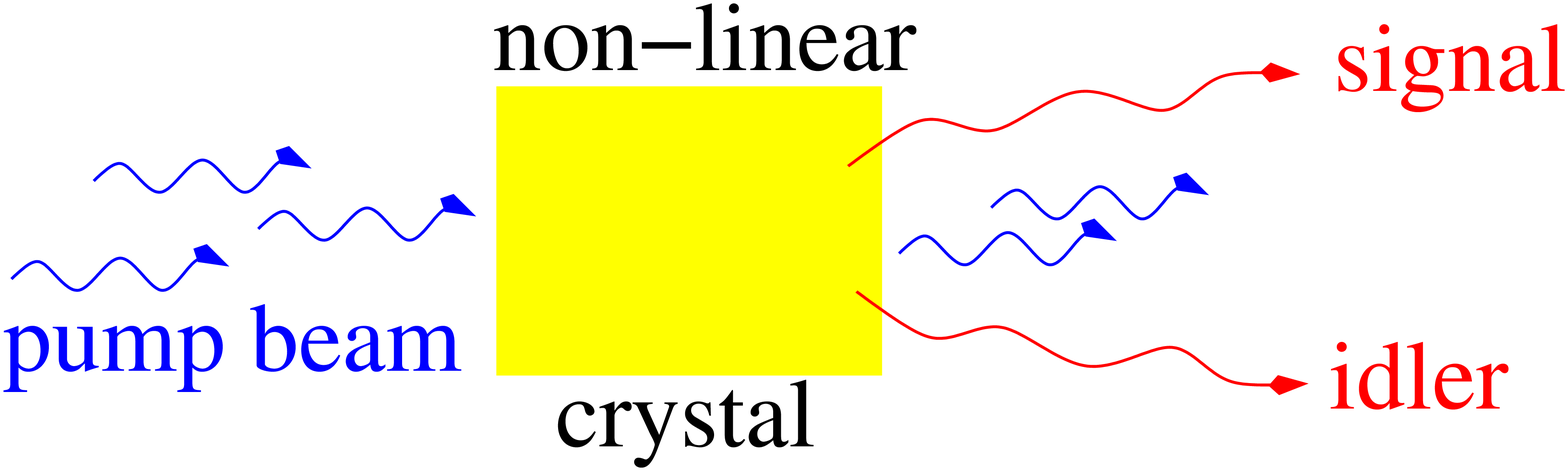,width=.5\textwidth}
\caption{Sketch of spontaneous parametric down-conversion.}
\label{down-conversion}
\end{figure}

Spontaneous parametric down-conversion occurs if a pump beam, 
e.g., a blue or UV laser, is sent into a non-linear Kerr medium.
Then a small fraction of the photons of the pump beam are converted 
into pairs of red or IR photons (called signal and idler) whose 
energies sum up to the energy of the original pump photon, 
see Fig.~\ref{down-conversion}. 

The signatures of the Unruh effect discussed above are very similar 
to the process of spontaneous parametric down-conversion:
The laser beam accelerating the electrons 
(as discussed in the previous Section) corresponds to the pump beam 
and the electrons represent the non-linear medium.
The created pairs of red or IR photons (called signal and idler) are 
completely analogous to the photon pairs representing the 
signatures of the Unruh effect.
In both cases, we have a spontaneous process driven by quantum 
fluctuations and the created photon pairs are entangled. 
However, the photon pairs representing the signatures of the Unruh 
effect can have much higher energies (e.g., in the keV or even MeV range) 
which might enable us to do quantum optics type experiments with 
entangled photon pairs in the keV-MeV regime\cite{Habs}. 

\section{Conclusions and Outlook}

By means of three examples -- the Schwinger mechanism, Hawking radiation, 
and the Unruh effect -- we studied fundamental quantum effects and 
established interesting analogies to quantum optics.
Even though the three physical scenarios are quite different, there are 
strong similarities:
Under extreme conditions, the quantum vacuum fluctuations can manisfest 
themselves as real particles.
These particles are always created in pairs 
(for bosons, we have a squeezed state).
All three phenomena are relativistic ($c$) quantum ($\hbar$) field effects. 
The analogy to laboratory physics (e.g., quantum optics) might help us 
to develop an experimental approach to these fundamental effects.  

\section*{Acknowledgments}

The author acknowledges support by the German Research Foundation (DFG)
under grant \# SCHU~1557/1 (Emmy-Noether Programme) and SFB-TR12. 


\end{document}